\documentclass[10pt,a4paper]{article}
\usepackage{amsmath}
\usepackage{graphicx}
\usepackage{amsfonts}
\usepackage{amssymb}
\usepackage[left=2cm,right=2cm,top=2cm,bottom=2cm]{geometry}
\usepackage{float}
\usepackage{bm}
\usepackage{color}
\usepackage{cite}

\usepackage[utf8]{inputenc}
\usepackage[english]{babel}
\usepackage{amsmath}
\usepackage{amsfonts}
\usepackage{amssymb}
\usepackage{float}
\usepackage[caption = false]{subfig}
\usepackage{graphicx}
\DeclareMathOperator\arctanh{arctanh}

\begin{document}

\title{EUP-corrected thermodynamics of BTZ black hole}
\author{ B. Hamil%
\thanks{%
hamilbilel@gmail.com} \\
D\'{e}partement de TC de SNV, Universit\'{e} Hassiba Benbouali, Chlef,
Algeria. \and B. C. L\"{u}tf\"{u}o\u{g}lu\thanks{%
bclutfuoglu@akdeniz.edu.tr (corresponding author)} \\
Department of Physics, Akdeniz University, Campus 07058, Antalya, Turkey, \\
Department of Physics, University of Hradec Kr\'{a}lov\'{e}, \\
Rokitansk\'{e}ho 62, 500 03 Hradec Kr\'{a}lov\'{e}, Czechia. \and
L. Dahbi\thanks{%
l.dahbi@ens-setif.dz} \\
Teacher Education College of Setif, Messaoud Zeghar, Algeria.}
\maketitle

\begin{abstract}
In this manuscript, we investigate the influence of the extended uncertainty principle (EUP) on the thermodynamics of a charged rotating Ba\~{n}ados, Teitelboim and Zanelli (BTZ) black hole in (2+1)-dimensional anti-de Sitter (AdS) and de Sitter (dS) space-time, respectively. We find that EUP-corrected Hawking temperature, entropy, volume, Gibbs free energy, pressure and heat capacity functions have different characteristic behaviours in (A)dS space-time. During the detailed analysis, we obtain a critical horizon in the AdS space-time where the pressure has a non-zero minimum value. However, in the dS space-time, we cannot achieve such a minimal pressure value.  Instead there is a "cut point" where the pressure becomes zero. We show that after that horizon value the black hole becomes unstable, with negative pressure which leads to the collapse of the black hole. 
\end{abstract}

\section{Introduction}
Entropy is one of the basic concepts of thermodynamics. A half-century ago, Bekenstein noticed an important similarity between entropy and black hole's area: both of them are increasing irreversibly \cite{Bekenstein1}. With that motivation, he presumed black holes as thermodynamical systems. He generalized the second law of thermodynamics and defined black holes' entropy in terms of their area \cite{Bekenstein2, Bekenstein3}. In the same year, Bardeen et al. gave the four laws of black hole mechanics and established their correspondences in thermodynamics \cite{Bardeen}. Soon after, Hawking showed that black holes emit radiation like black bodies according to a well-defined characteristic  temperature, which is known as the Hawking temperature in literature \cite{Hawking1, Hawking2}. Since then, many research are done to demonstrate the correlation between black hole thermodynamics and real systems\cite{Mann1, Tzikas, Wang}. We refer two important review papers \cite{Waldrev, Carliprev} and the references therein,   that summarize its history and the latest developments on the field.

On the atomic scale, according to Heisenberg’s uncertainty principle, there is a basic limit on measuring the momentum and position simultaneously. Actually, this principle sets a conditional minimum if the other one is fixed. For this reason, it is said that Heisenberg's principle does not impose an absolute minimum or maximum value on momentum and position. Therefore, it is believed that Heisenberg's principle needs some modification especially when the non-negligible gravitational effects are taken into account \cite{Wigner1957, Wigner1958, Amati1989}. Kempf et al. proposed a modification known in the literature as the generalized uncertainty principle (GUP), which defines the minimum length. \cite{Kempf1995}. Soon after it is understood that in the GUP scenario black hole thermodynamics is also modified, especially in the following aspects. First, the Hawking temperature can change in a way that precludes the black hole from evaporating \cite{Adler2001, Nozari2005}. Second, the entropy-area relation alters because of the Hawking-entropy modification \cite{Medved2004, Zhao2006}. However, in the large length scales, like in anti-de-Sitter (AdS) and de Sitter (dS) backgrounds, GUP modifications should not be expected to be important. With this motivation, Park used the extended uncertainty principle (EUP), which was proposed by Bolen and Cavaglia to consider the large length scale corrections \cite{Bolen2005},  to discuss the Hawking temperature in (A)dS space-time  \cite{Park2008}. Since Heisenberg uncertainty principle modification is not unique, many research on various black hole's thermodynamics are done by considering different deformation scenarios \cite{Ong2018, hha1, cc14, cc15, cc17, cc19, cc20, cc21}.

BTZ black hole solution is first derived by Ba\~nados, Teitelboim, and Zanelli while they were examining  the vacuum solution of the Einstein-Maxwell equation  with a negative cosmological constant in (2+1) dimensions \cite{BTZ}. They showed that the BTZ black hole with event and inner horizons is well defined in terms of charge, angular momentum, and mass. Unlike the Schwarzschild and Kerr black holes, the BTZ black hole has a constant curvature which is not singular at origin \cite{Carlip1995}. For this reason, it has attracted attention and many studies have been conducted on it \cite{rcBTZ, Wu2006, Rahaman2013, Anacleto2015, Hendi2016, Hendi2017, Sadeghi2017, Anacleto2018, Panah2019, Singh2020, babar, Kazempour2020, Anacleto2021, BCL2022}. For example, Ach\'ucarro and Ortiz revisited the BTZ solutions with dimensional reduction techniques \cite{rcBTZ}. Wu and Jiang showed that entropy-area formulae would be consistent if one considers the Hawking radiation as a semi-classical tunneling process \cite{Wu2006}.  Rahaman et al. derived the BTZ black hole solution out of noncommutative geometry \cite{Rahaman2013}. Anacleto et al. considered the GUP formalism to obtain quantum corrections of the BTZ black hole's entropy in noncommutative space-time \cite{Anacleto2018}, just a year later Sadeghi's work in the commutative space-time \cite{Sadeghi2017}. In 2020, Singh et al. investigated the Hawking radiation of BTZ black hole in the presence of GUP deformation  \cite{Singh2020}. In the same year, Babar et al. handled the similar problem for the BTZ-like black hole that is surrounded by the quintessence matter \cite{babar}.  Very recently, the authors of this manuscript explored the thermodynamic properties of BTZ black hole in the gravity's rainbow context \cite{BCL2022}. 

Since the GUP formalism influences the thermodynamics of the BTZ black hole drastically, we find extremely interesting to investigate the effect of the EUP formalism on the same type of black holes in (A)dS space-time. We organize the manuscript as follows: In Sec. \ref{sec2}, we introduce the considered  charged rotating BTZ black hole and EUP formalism in (A)dS space-time. In Sec. \ref{sec3}, we examine and compare the  Hawking temperature, entropy, thermodynamical volume, Gibbs free energy, pressure, and heat capacity functions of the black hole. Finally, we conclude the manuscript with a brief section.

\section{Charged rotating BTZ black hole and EUP formalism} \label{sec2}
We consider a charged rotating BTZ black hole \cite{BTZ}
\begin{equation}
ds^{2}=-\mathcal{F}\left( r\right) dt^{2}+\frac{1}{\mathcal{F}\left(
r\right) }dr^{2}+r^{2}\left( d\phi -\frac{J}{2r^{2}}dt\right)^2 ,
\end{equation}
with the following lapse function
\begin{equation}
\mathcal{F}\left( r\right) =-M+\frac{r^{2}}{\ell ^{2}}+\frac{J^{2}}{4r^{2}}%
-2Q^{2}\ln \frac{r}{\ell }.
\end{equation}
Here,  $\ell$ denotes the (A)dS radius, while $Q$, $J$, and  $M$ indicate the black hole's charge, angular momentum (spin) and mass, respectively. As discussed in \cite{2},  the lapse function does not have roots at some parameter values. For example, a BTZ black hole with a sufficiently small mass would give rise to a naked singularity, since there is no event horizon if all other parameters are equal to one \cite{Hendi}. If the BTZ black holes are massive enough, then they have outer and inner event horizons, which are also known as Killing and Cauchy horizons in literature. 

In this manuscript, we examine the thermodynamics of the BTZ black hole in the Mignemi's EUP formalism  \cite{Mignemi2009}. According to him,  Heisenberg uncertainty relation could be modified by 
\begin{eqnarray}
&&\Delta P\Delta X\geq \left( 1+\alpha \left( \Delta X\right) ^{2}\right) , \quad \alpha>0. \label{AdSEUP} 
\end{eqnarray}
in the AdS space-time, and
\begin{eqnarray}
&&\Delta P\Delta X\geq \left( 1-\big|\alpha\big| \left( \Delta X\right) ^{2}\right) , \quad \alpha<0. \label{dSEUP}
\end{eqnarray}
in the dS space-time. Here, $\alpha$ is proportional to the cosmological constant. Alternatively, the deformation parameter can be given in terms of a dimensionless parameter, $\alpha_0$, and space-time radius as:
\begin{eqnarray}
\alpha&=& \frac{\alpha_0}{\ell^2},
\end{eqnarray}
where $\ell^2>0$ and $\ell^2 < 0$ in the AdS and dS space-times, respectively.

\section{EUP-corrected thermodynamics}\label{sec3}

We start by using the conventional definition of the Hawking temperature \cite{chen1}
\begin{eqnarray}
T_H=\frac{\kappa }{8\pi }\frac{dA}{dS},
\end{eqnarray}
where $\kappa $, the surface gravity at the outer horizon,  is equal to
\begin{equation}
\kappa =\left. \frac{\partial \mathcal{F}\left( r\right) }{\partial r}\right
\vert _{r=r_{+}}=\frac{2r_{+}}{\ell^{2}}-\frac{J^{2}}{2r_{+}^{3}}-\frac{2Q^{2}}{r_{+}}.
\end{equation}%
According to the heuristics \cite{xiang}, when a particle near the event horizon is absorbed by a black hole, the black hole's area and entropy increase. This enhancement can be taken at least  as
\begin{equation}
\frac{\left( \Delta A\right) _{\min }}{\left( \Delta S\right) _{\min }}=%
\frac{\gamma }{\ln 2}\left( \Delta X\right) \left( \Delta P\right). 
\end{equation}
where $\gamma$ is a calibration factor, that helps to match the result with the ordinary result. In the  EUP scenario, we express the EUP-corrected Hawking temperature by
\begin{equation}
T_{EUP}=\frac{\kappa }{8\pi }\frac{\gamma }{\ln 2}\left( \Delta X\right) \left( \Delta P\right).
\end{equation}
We model the BTZ black hole as a black box with linear size $r_{+}$. Then, we take the
uncertainty in the position with the radius of the event horizon as $\Delta X =r_{+}$. For $\gamma=2 \ln 2$, we obtain the Hawking temperature in the form of
\begin{eqnarray} \label{TempHaw}
T_{H-EUP}=\left \{
\begin{array}{ll}
\frac{1}{4\pi }\left( \frac{2r_{+}}{\ell ^{2}}-\frac{J^{2}}{2r_{+}^{3}}-%
\frac{2Q^{2}}{r_{+}}\right) \Big( 1+\alpha r_{+}^{2}\Big),  &\alpha >0 , \\
\frac{1}{4\pi }\left( \frac{2r_{+}}{\ell ^{2}}-\frac{J^{2}}{2r_{+}^{3}}-%
\frac{2Q^{2}}{r_{+}}\right) \Big( 1-\left \vert \alpha \right \vert
r_{+}^{2}\Big), &\alpha <0 .%
\end{array}%
\right. 
\end{eqnarray}
For $\alpha=0$, Eq. \eqref{TempHaw} gives its standard form. The Hawking  temperature of a black hole must be positive-valued, therefore in both space-time  we have the following constraint on the event horizon radius.  
\begin{eqnarray}
r_{+} \geq \frac{Q \ell}
{\sqrt{2}}\sqrt{1+\sqrt{1+\frac{J^2}{Q^4 \ell^2}}}.
\end{eqnarray}
However, in the dS background an additional constraint arises that limits the radius value from above
\begin{equation}
r_{+}\leq \frac{1}{\sqrt{\left \vert \alpha \right \vert }}.
\end{equation}
Therefore, we conclude that the EUP-corrected Hawking temperature is physical only in the following intervals:
\begin{eqnarray} \label{Tempdef}
\begin{array}{rllll}
r_{+} \geq \frac{Q \ell}
{\sqrt{2}}\sqrt{1+\sqrt{1+\frac{J^2}{Q^4 \ell^2}}}. & &\alpha >0 , \\
\frac{1}{\sqrt{\left \vert \alpha \right \vert }} \geq r_{+} \geq \frac{Q \ell}{\sqrt{2}}\sqrt{1+\sqrt{1+\frac{J^2}{Q^4 \ell^2}}} .
& &\alpha <0 .%
\end{array}%
\end{eqnarray}
In the rest of the manuscript, we take $Q=0.05$, $J=1$, and $\ell=4$. Therefore, we find the minimal physical horizon value as $1.42$.

In Fig. 1, we depict the EUP-corrected Hawking temperature in AdS and dS backgrounds. For comparison, we consider the HUP limit in addition to three different deformation parameters.

\begin{figure}[htb]\label{fig:Hawking}
   \begin{minipage}[t]{.45\textwidth}
        \centering
        \includegraphics[width=\textwidth]{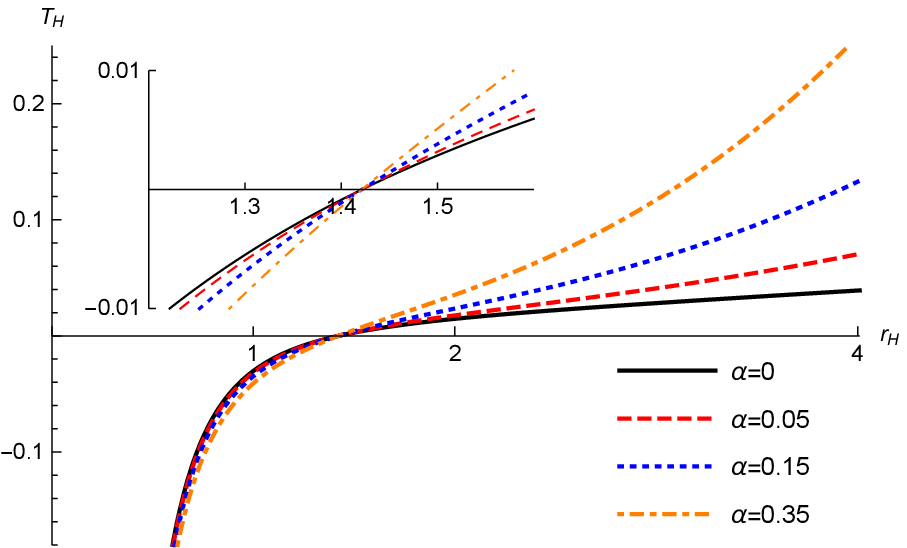}
    \end{minipage}
    \hfill
    \begin{minipage}[t]{.45\textwidth}
        \centering
        \includegraphics[width=\textwidth]{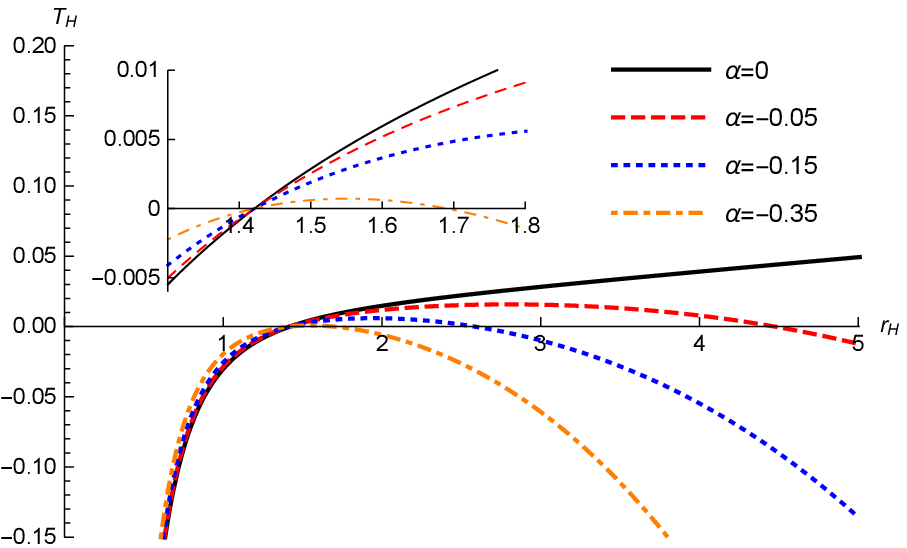}
    \end{minipage}  
    \caption{EUP corrected Hawking temperature versus horizon in AdS and dS space-time.}
\end{figure}
We observe that in the AdS space-time, EUP corrections increase the Hawking temperature in the physical region. In the dS space-time,  at the end of the evaporation process, the EUP-corrected Hawking temperature goes to zero at $4.47$, $2.58$, and $1.69$ for the following deformation parameters $0.05$, $0.15$, and $0.35$, respectively, as predicted in Eq. \eqref{Tempdef}.  

Next, we derive the EUP-corrected entropy function by utilizing 
\begin{equation}
S=\int \frac{dM}{T}.
\end{equation}
We find
\begin{equation}\label{enteup}
S_{EUP}=\left\{
\begin{array}{ll}
\frac{4\pi }{\sqrt{\alpha }}\arctan \left( \sqrt{\alpha }%
r_{+}\right) ,  & \alpha >0 \\
\frac{4\pi }{\sqrt{\left\vert \alpha \right\vert }}\arctanh\left(
\sqrt{\left\vert \alpha \right\vert }r_{+}\right) , &
\alpha <0%
\end{array}%
\right. .
\end{equation}
For $\alpha=0$, Eq. \eqref{enteup} reduces to its usual form, $4\pi r_+$. By expanding $S_{EUP}$ in its power series over $\alpha $, we get%
\begin{equation}
S_{EUP}=\left\{
\begin{array}{ll}
 4\pi r_{+}-\frac{4\pi }{3}\alpha r_{+}^{3}, & \alpha>0 \\
4\pi r_{+}+\frac{4\pi }{3}\left\vert \alpha \right\vert r_{+}^{3}, & \alpha <0%
\end{array}%
\right. .
\end{equation}

\begin{figure}[htbp]
   \begin{minipage}[t]{.45\textwidth}
        \centering
        \includegraphics[width=\textwidth]{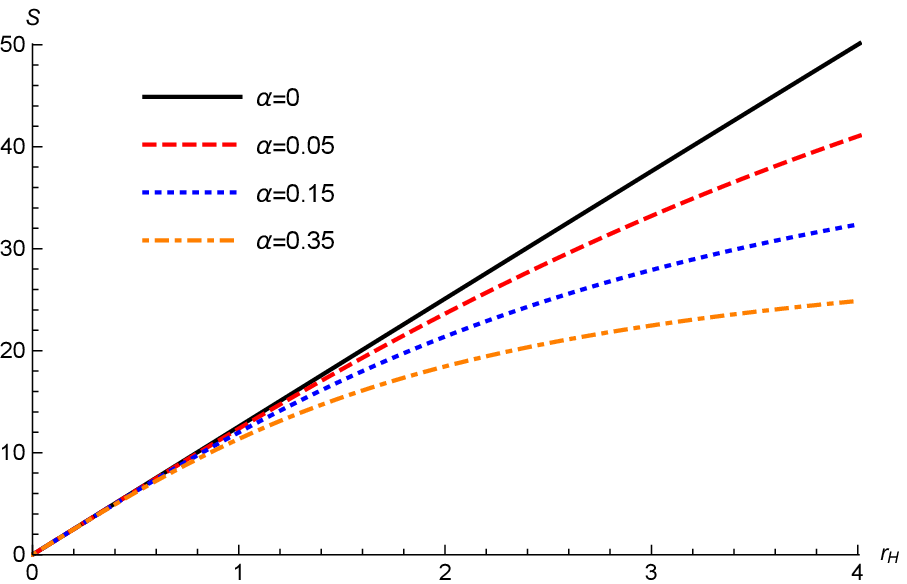}
    \end{minipage}
    \hfill
    \begin{minipage}[t]{.45\textwidth}
        \centering
        \includegraphics[width=\textwidth]{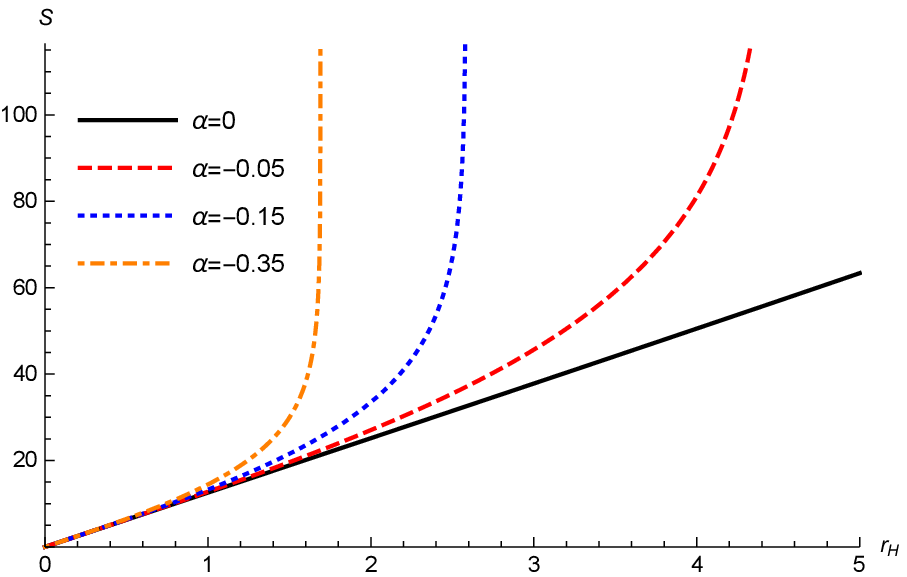}
    \end{minipage}  
    \label{fig:entropy}
    \caption{EUP corrected entropy versus horizon in AdS and dS space-time.}
\end{figure}
We observe that, in the AdS space, the correction function tends to decrease the BTZ black hole entropy, while in dS space, the correction function tends to increase the BTZ black hole entropy. In Fig. 2, we depict the black hole entropy versus the horizon radius in the HUP and EUP approaches, respectively. In the dS space-time, the black hole entropy goes to infinity at $%
r_{+}=r_{\max }=\frac{1}{\sqrt{\left\vert \alpha \right\vert }}$. 
The entropy function in the case of AdS space, however, gets its maximum value  when $r_{+}$ goes to infinity.

\begin{figure}[htbp]
   \begin{minipage}[t]{.45\textwidth}
        \centering
        \includegraphics[width=\textwidth]{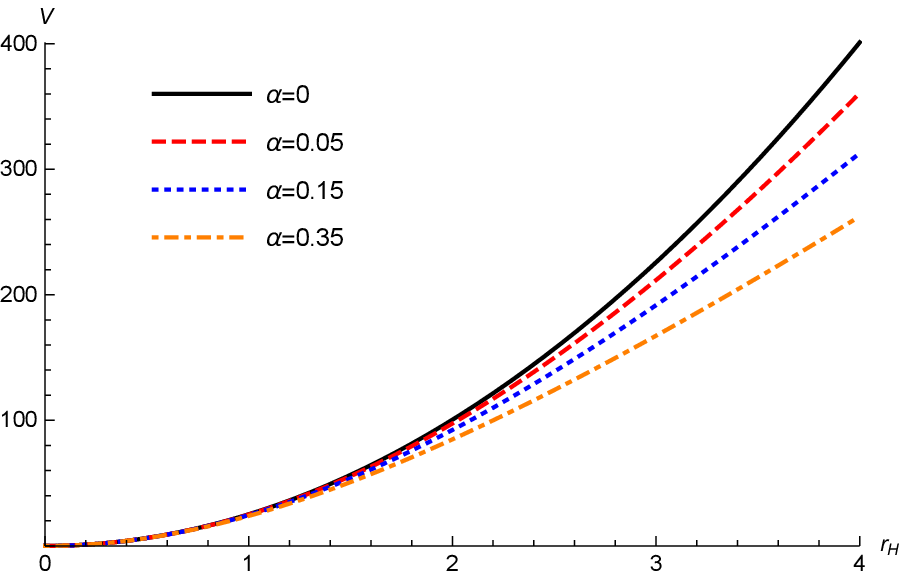}
    \end{minipage}
    \hfill
    \begin{minipage}[t]{.45\textwidth}
        \centering
        \includegraphics[width=\textwidth]{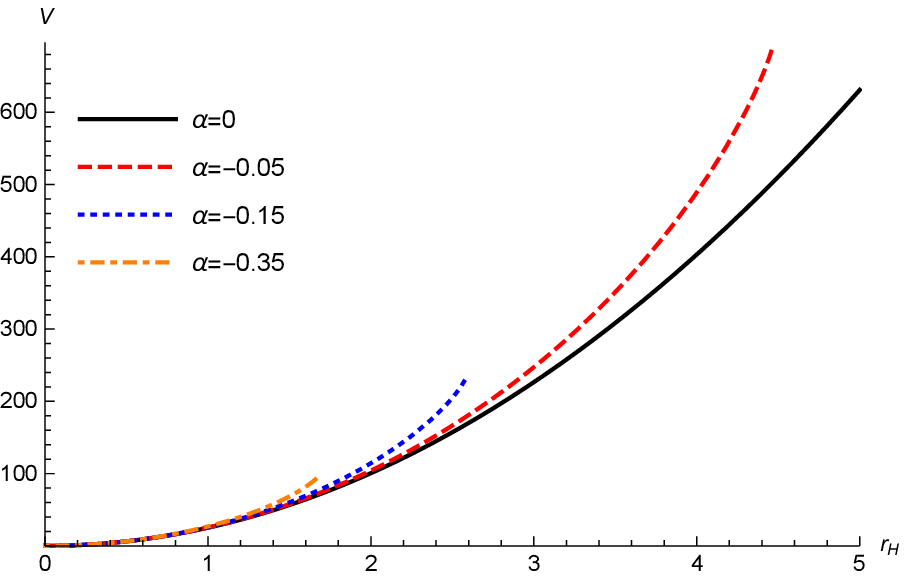}
    \end{minipage}  
    \label{fig3:volume}
    \caption{EUP corrected volume versus horizon in AdS and dS space-time.}
\end{figure}

Next, we investigate the EUP-corrected volume function. We use
\begin{equation}
V=4\int Sdr_{+},
\end{equation}%
to derive the thermodynamic volume. We find
\begin{equation} \label{voleup}
V_{EUP}=\left\{ 
\begin{array}{ll}
\frac{16\pi }{\alpha }\bigg[ \sqrt{\alpha } r_{+}\arctan \left( \sqrt{\alpha }%
r_{+}\right) -\frac{1 }{2 }\log \left( 1+\alpha r_{+}^{2}\right) \bigg] , & \alpha >0 \\ 
\frac{16\pi }{\left\vert \alpha \right\vert }\bigg[  \sqrt{\left\vert \alpha \right\vert}r_{+}\arctanh%
\left( \sqrt{\left\vert \alpha \right\vert }r_{+}\right) +\frac{1 }{2}\log \left( 1-\left\vert \alpha
\right\vert r_{+}^{2}\right) \bigg], & \alpha <0%
\end{array}%
\right. .
\end{equation}
For $\alpha=0$, we note that Eq. \eqref{voleup} reduces to $8\pi r_+^2$. By expanding $V_{EUP}$ in power series over the deformation parameter, we get%
\begin{equation}
V_{EUP}=\left\{
\begin{array}{ll}
 8\pi r_{+}^2-\frac{4\pi }{3}\alpha r_{+}^{4}, & \alpha>0 \\
8\pi r_{+}^2+\frac{4\pi }{3}\left\vert \alpha \right\vert r_{+}^{4}, & \alpha <0%
\end{array}%
\right. .
\end{equation}
We depict the EUP-corrected volume versus horizon in Fig. \eqref{fig3:volume}. 

We observe that EUP corrections decrease the volume in AdS space-time. In the dS background, EUP-corrections set an upper bound to volume at $r_{max}$. 

\newpage

Then, we derive the thermodynamical formula of the Gibbs free energy 
\begin{equation}
G=M-TS.
\end{equation}%
We find,
\begin{equation}
G=\left\{
\begin{array}{ll}
\frac{r_{+}^{2}}{\ell ^{2}}+\frac{J^{2}}{4r_{+}^{2}}-2Q^{2}\ln \frac{r_{+}}{%
\ell }-\frac{\left( 1+\alpha r_{+}^{2}\right) }{\sqrt{\alpha }}%
\left( \frac{2r_{+}}{\ell ^{2}}-\frac{J^{2}}{2r_{+}^{3}}-\frac{2Q^{2}}{r_{+}}%
\right) \arctan \left( \sqrt{\alpha }r_{+}\right) , &
\alpha >0 \\
\frac{r_{+}^{2}}{\ell ^{2}}+\frac{J^{2}}{4r_{+}^{2}}-2Q^{2}\ln \frac{r_{+}}{%
\ell }-\frac{\left( 1-\left\vert \alpha \right\vert r_{+}^{2}\right) }{%
\sqrt{\left\vert \alpha \right\vert }}\left( \frac{2r_{+}}{\ell ^{2}}-%
\frac{J^{2}}{2r_{+}^{3}}-\frac{2Q^{2}}{r_{+}}\right) \arctanh\left(
\sqrt{\left\vert \alpha \right\vert }r_{+}\right) , &
\alpha <0%
\end{array}%
\right. .
\end{equation}
which reduces to 
\begin{eqnarray}
G=-\frac{r_{+}^{2}}{\ell^{2}}+\frac{3J^{2}}{4r_{+}^{2}}+2Q^{2}\bigg(1-\ln \frac{r_{+}}{\ell } \bigg),
\end{eqnarray}
for $\alpha=0$. We present the plot of the EUP-corrected Gibbs free energy versus horizon in Fig. (4). We observe that in the AdS space-time, the EUP corrected Gibbs free energy takes lower values for larger deformation parameters in the physical region. However in the dS space-time, it exhibits a different character. We see a maximal value at   $r_+=\frac{1}{\sqrt{\left\vert \alpha \right\vert }}$.

\begin{figure}[htbp]
   \begin{minipage}[t]{.45\textwidth}
        \centering
        \includegraphics[width=\textwidth]{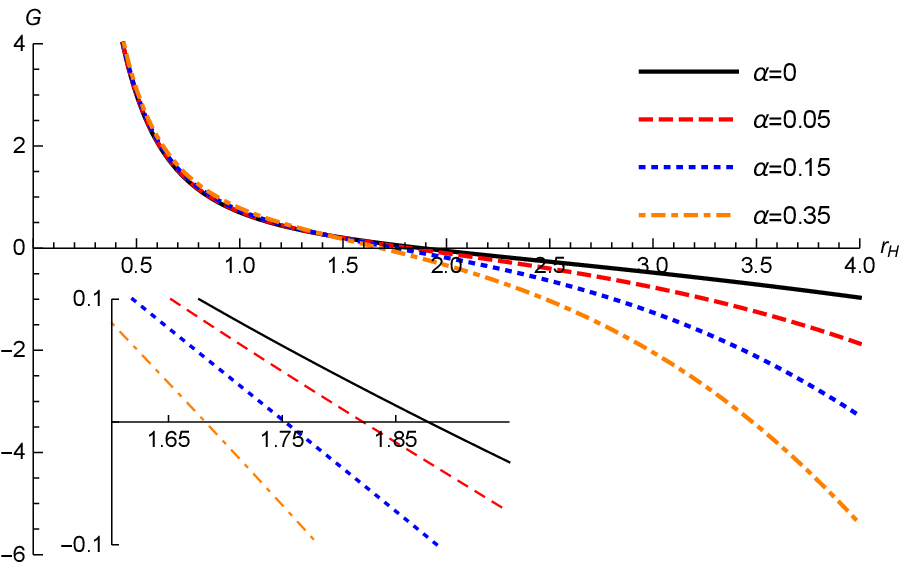}
    \end{minipage}
    \hfill
    \begin{minipage}[t]{.45\textwidth}
        \centering
        \includegraphics[width=\textwidth]{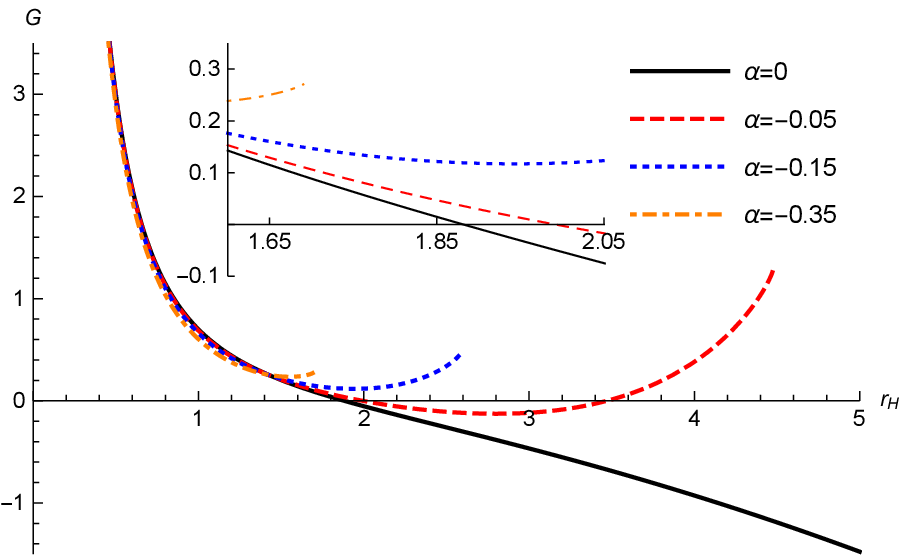}
    \end{minipage}  
    \label{fig:gibbs}
    \caption{EUP corrected Gibbs free energy versus horizon in AdS and dS space-time.}
\end{figure}
Next, we obtain the EUP-corrected pressure via 
\begin{eqnarray}
P&=&\frac{\partial G}{\partial V}.
\end{eqnarray}
We find 
\begin{equation}
P_{EUP}=\left\{ 
\begin{array}{ll}
\frac{1}{16\pi }\bigg[\left( \frac{2}{\ell ^{2}}+\frac{3J^{2}}{2r_{+}^{4}}+\frac{%
2Q^{2}}{r_{+}^{2}}\right) +\alpha r_{+}^{2}\left( \frac{6}{\ell ^{2}}+\frac{%
J^{2}}{2r_{+}^{4}}-\frac{2Q^{2}}{r_{+}^{2}}\right) \bigg] , & \alpha >0 \\ 
\frac{1}{16\pi }\bigg[\left( \frac{2}{\ell ^{2}}+\frac{3J^{2}}{2r_{+}^{4}}+\frac{%
2Q^{2}}{r_{+}^{2}}\right) -\left\vert \alpha
\right\vert r_{+}^{2}\left( \frac{6}{\ell ^{2}}+\frac{%
J^{2}}{2r_{+}^{4}}-\frac{2Q^{2}}{r_{+}^{2}}\right)\bigg] , &\alpha <0%
\end{array}%
\right. .
\end{equation}
For $\alpha=0$, pressure reads
\begin{eqnarray}
P=\frac{1}{8\pi }\left( \frac{1}{\ell ^{2}}+\frac{3J^{2}}{4r_{+}^{4}}+\frac{%
Q^{2}}{r_{+}^{2}}\right).
\end{eqnarray}
We illustrate the EUP-corrected pressure function in Fig. (5). 

\begin{figure}[htbp]
   \begin{minipage}[t]{.45\textwidth}
        \centering
        \includegraphics[width=\textwidth]{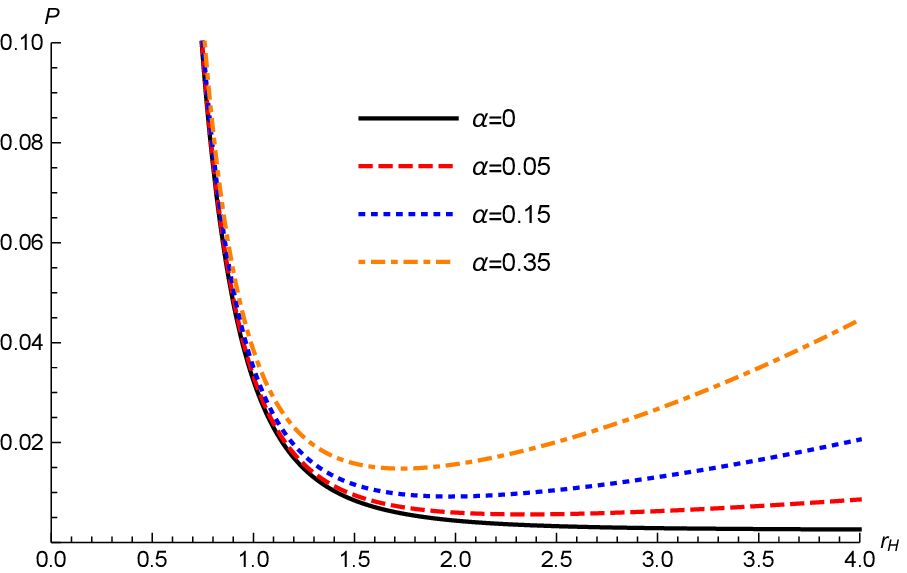}
    \end{minipage}
    \hfill
    \begin{minipage}[t]{.45\textwidth}
        \centering
        \includegraphics[width=\textwidth]{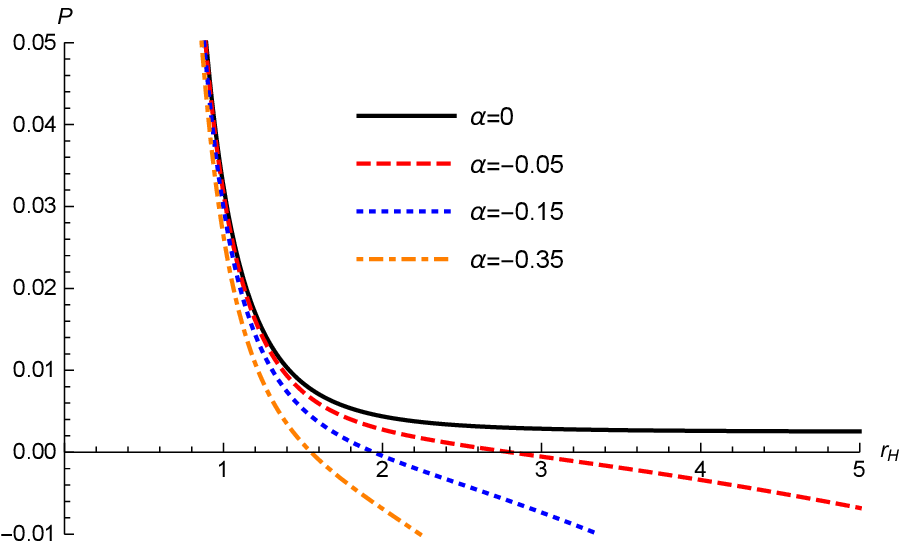}
    \end{minipage}  
    \label{fig:pressure}
    \caption{EUP corrected pressure versus horizon in AdS and dS space-time.}
\end{figure}

In AdS space-time, we observe that the pressure starts to increase after critical horizon value.These critical horizon values, $r_{+_{critic}}$, can be found by setting the first derivative of the EUP-corrected pressure equal to zero. 
\begin{eqnarray}
-\frac{1}{4\pi}\bigg(\frac{Q^2}{ r_+^3}+\frac{3 J^2}{2 r_+^5}\bigg)+ \frac{\alpha}{4\pi}\bigg(\frac{3 r_+}{\ell^2}-\frac{J^2}{4 r_+^3}\bigg)=0
\end{eqnarray}
We observe that for $\alpha=0$, there is no real critical horizon value. In AdS space-time, according to the selected parameters, we obtain only one real and positive critical horizon value as given in Table 1.

\begin{table}[ht]
\caption{Critical horizon in AdS space-time} 
\centering 
\begin{tabular}{c c c c} 
\hline\hline 
$\alpha$ &$r_{+_{critic}}$ & $P_{min}$ & V \\ [0.5ex] 
\hline 
0.05 & 2.35  &0.0056 & 133.13  \\ 
0.15 & 1.97  &0.0092 & 89.97  \\
0.35 & 1.73  &0.0148 & 65.81  \\   [1ex] 
\hline 
\end{tabular}
\label{table:nonlin1} 
\end{table}

In dS space-time, in the physical region the pressure decreases faster for greater deformation parameter value as shown in Fig. (5). On contrary to AdS space-time, the pressure does not have a minimal value. At $r_{+_{cut}}$ value, pressure becomes zero. Then, we observe negative pressure. Although  the volume increase between $r_{+_{cut}}< r_+ <r_{max} $, the negative pressure increases, and at the certain point black hole collapses (or we lose information). We tabulate this scenario in Table 2.  

\begin{table}[ht]
\caption{Pressure-volume and horizon correlation in dS space-time} 
\centering 
\begin{tabular}{c c c c c c c c c c} 
\hline\hline 
$\alpha$ & $r_{phys}$ &  $P_{phys}$ & $V_{phys}$& $r_{+_{cut}}$ & $P_{cut}$ & $V_{cut}$ &$r_{max}$ & $P_{min}$ & $V_{min}$ \\ [0.5ex] 
\hline 
0.05 & 1.42  &0.0089 & 51.67  & 2.80 & 0 & 212.62 & 4.47 & -0.0049 & 696.83  \\ 
0.15 & 1.42  &0.0069 & 53.71  & 1.94 & 0 & 106.87 & 2.58 & -0.0045 & 232.52  \\
0.35 & 1.42  &0.0029 & 59.66  & 1.55 & 0 & 74.82 & 1.69 & -0.0024 &  100.22 \\   [1ex] 
\hline 
\end{tabular}
\label{table:nonlin2} 
\end{table}

Finally we examine the EUP-corrected heat capacity of the BTZ black hole by using the following relation 
\begin{eqnarray}
C_{J,Q}=T_{H}\left( \frac{\partial S}{\partial T_{H}}\right) 
_{J,Q}
\end{eqnarray}
We get
\begin{equation} \label{heatfunceup}
C_{J,Q-EUP}=\left\{
\begin{array}{ll}
4\pi r_{+} \bigg(\frac{2}{\ell ^{2}}-\frac{J^{2}}{2r_{+}^{4}}-\frac{2Q^{2}}{
r_{+}^2}\bigg) \bigg[\left( \frac{2}{\ell ^{2}}+\frac{3J^{2}}{2r_{+}^{4}}+\frac{2Q^{2}}{%
r_{+}^{2}}\right) +\alpha r_{+}^{2}\left( \frac{6}{\ell ^{2}}+\frac{J^{2}%
}{2r_{+}^{4}}-\frac{2Q^{2}}{r_{+}^{2}}\right)\bigg]^{-1}, &\alpha
>0 \\
4\pi r_{+} \bigg(\frac{2}{\ell ^{2}}-\frac{J^{2}}{2r_{+}^{4}}-\frac{2Q^{2}}{
r_{+}^2}\bigg)\bigg[\left( \frac{2}{\ell ^{2}}+\frac{3J^{2}}{2r_{+}^{4}}+\frac{2Q^{2}}{%
r_{+}^{2}}\right) -\left\vert \alpha \right\vert r_{+}^{2}\left( \frac{6%
}{\ell ^{2}}+\frac{J^{2}}{2r_{+}^{4}}-\frac{2Q^{2}}{r_{+}^{2}}\right)\bigg]^{-1}, 
&\alpha <0%
\end{array}%
\right. .
\end{equation}
which reduces to 
\begin{equation} 
C_{J,Q}= - (4\pi r_{+}) \frac{J^2 \ell^2 + 4 r_+^2 (\ell^2 Q^2 - r_+^2)}{3J^2 \ell^2 + 4 r_+^2 (\ell^2 Q^2 + r_+^2)}.
\end{equation}
for $\alpha=0$. We note that there is no singularity in this case. Then, we plot the EUP-corrected specific heat function versus horizon in Fig. (6). 

\begin{figure}[htbp]
   \begin{minipage}[t]{.45\textwidth}
        \centering
        \includegraphics[width=\textwidth]{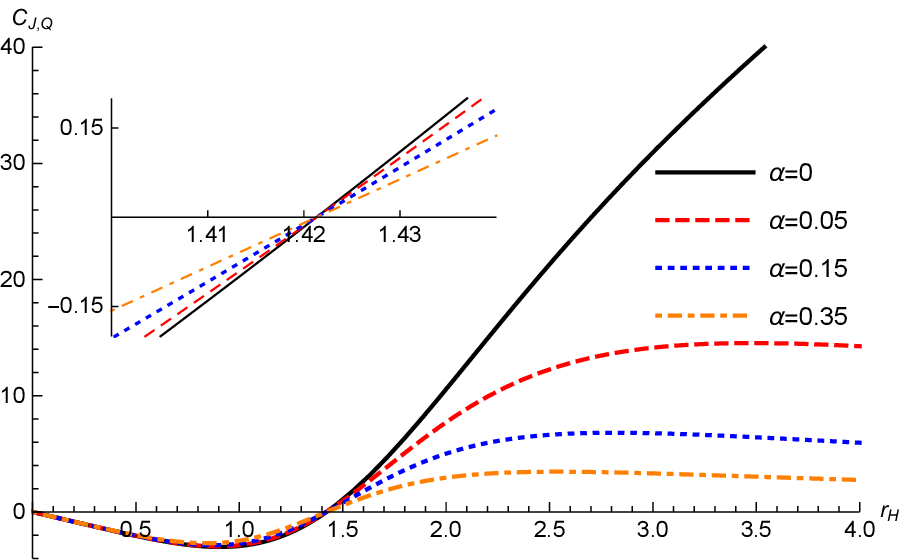}
    \end{minipage}
    \hfill
    \begin{minipage}[t]{.45\textwidth}
        \centering
        \includegraphics[width=\textwidth]{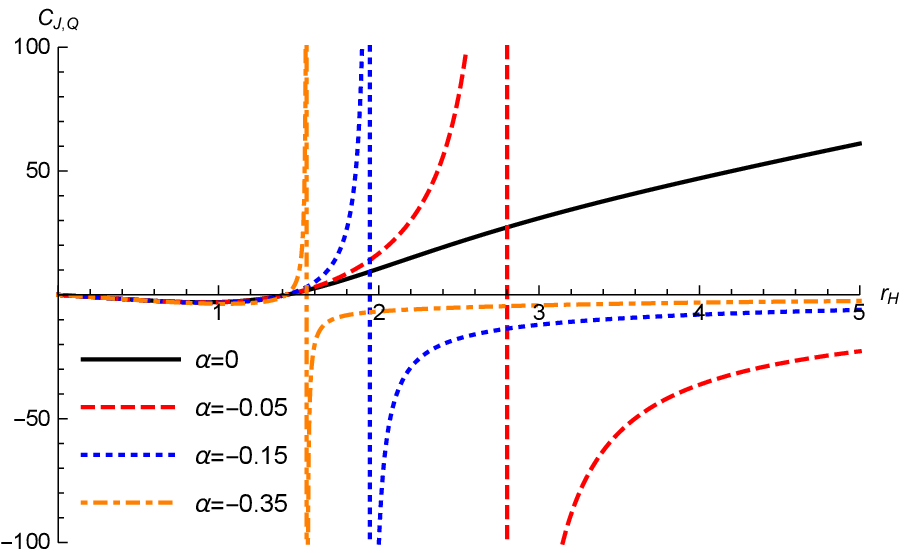}
    \end{minipage}  
    \label{fig:specificheat}
    \caption{EUP corrected specific heat versus horizon in AdS and dS space-time.}
\end{figure}
In AdS space-time we see that the EUP-corrected BTZ black hole is stable only in the physical region. On the other hand, we observe that black hole is also stable in $r_{Phys}< r_+ < r_{cut}$ in dS space-time, where the pressure takes only positive values. 
\section{Conclusion}

In this manuscript, we consider a charged rotating Ba\~{n}ados, Teitelboim and Zanelli (BTZ) black hole in (2+1)-dimensional (anti)-de Sitter (A)dS space-time, and accordingly, we examine the quantum corrections added to its thermal quantities in the extended uncertainty principle (EUP) formalism. At first, we find that the event horizon has an extra upper bound only in dS space-time in addition to the physical lower bound that exists in both space-times. In the physical range of the AdS space-time, we observe that the Hawking temperature increases with the quantum corrections. However, in the dS space-time, we show that the EUP-corrected Hawking temperature takes positive values only in an interval. Then we examine the entropy function. We find that quantum corrections decrease the usual entropy value in AdS space-time. However in dS space-time, entropy goes to infinity at the upper horizon limit. When we examine the volume, we saw a decrease in AdS space-time. On contrary, in dS space-time, the volume becomes greater with the quantum corrections. In order to study the pressure, we first derive the EUP-corrected Gibbs free energy function. We detect different characteristics behaviour in (A)dS space-time. Next, we obtain the EUP-corrected pressure functions. We observe that the pressure in AdS space-time has a positive non-zero minimal value in the physical region. We derive an analytical expression to evaluate this critical horizon value. In dS space-time, we see that there is not any critical value as in AdS space-time, however, there is cut horizon value where the pressure become zero. We see that after this cut horizon value, the pressure becomes negative until the black hole collapses. We give detailed calculation for randomly chosen black hole parameters. Finally, we examine the specific heat function. We find that in AdS space-time, the black hole is stable in the physical range. In dS space-time, BTZ black hole is stable only between the physical and cut horizon values.

\section*{Acknowledgments}
One of the authors of this manuscript, BCL, is supported by the Internal  Project,  [2022/2218],  of  Excellent  Research  of  the  Faculty  of  Science  of Hradec Kr\'alov\'e University.

\end{document}